\documentclass[10pt,conference]{IEEEtran}
\IEEEoverridecommandlockouts
\usepackage{cite}
\usepackage{amsmath,amssymb,amsfonts}
\usepackage{algorithmic}
\usepackage{graphicx}
\usepackage{textcomp}
\usepackage{xcolor}
\usepackage{url}
\usepackage{subfig}
\usepackage{balance}
\usepackage{placeins}
\usepackage{fancyhdr}

\fancypagestyle{firstpage}
{
  \fancyhf{}
  
  \fancyfoot[C]{\footnotesize \copyright 2025 IEEE. Personal use of this material is permitted. Permission from IEEE must be obtained for all other uses, in any current or future media, including reprinting/republishing this material for advertising or promotional purposes, creating new collective works, for resale or redistribution to servers or lists, or reuse of any copyrighted component of this work in other works.}
}

\def\BibTeX{{\rm B\kern-.05em{\sc i\kern-.025em b}\kern-.08em
    T\kern-.1667em\lower.7ex\hbox{E}\kern-.125emX}}

\begin{document}

\title{Adapting to Reality: Over-the-Air Validation of AI-Based Receivers Trained with Simulated Channels}

\author{\IEEEauthorblockN{Riku Luostari}
    \IEEEauthorblockA{
        \textit{Nokia Mobile Networks,} \\
        Espoo, Finland \\
    }

    \and
    \IEEEauthorblockN{Dani Korpi, Mikko Honkala, Janne~M.J. Huttunen}
    \IEEEauthorblockA{\textit{Nokia Bell Labs} \\
        Espoo, Finland \\
    }
}
\maketitle

\begin{abstract}

    Recent research shows that integrating artificial intelligence (AI) into wireless communication systems can significantly improve spectral efficiency. However, most AI-based receiver studies rely on simulated radio channel data for both training and validation, raising concerns about real-world generalization, which is vital for ensuring reliable field performance. In this study, we train DeepRx, a convolutional neural network (CNN)-based OFDM receiver, under various simulated channel scenarios and validate its performance over-the-air (OTA) using software-defined radio (SDR) technology in a small cell-type setup. To enhance receiver training, we investigate a randomized 3GPP TS38.901 channel model to diversify the training data, thereby improving performance over conventional receivers and matching or exceeding the performance of receivers trained on narrowly targeted channel models. These results demonstrate DeepRx’s robust generalization capability and suggest that narrowly scoped, individual TS38.901 models can compromise both training and validation, underscoring the need for tailored channel models, careful training strategies, and OTA testing in learned receiver development.

\end{abstract}
\thispagestyle{firstpage}

\section{Introduction}

By integrating artificial intelligence (AI) into the physical layer of radio receivers, wireless communication systems can achieve significant gains in spectral efficiency compared to conventional heuristic methods \cite{osheaIntroductionDeepLearning2017a, huangDeepLearningPhysicalLayer2020a}. However, these learned receivers demand extensive data for effective training, and radio channel simulations can serve as a practical source of such data. Although multiple channel-modeling techniques exist, such as ray tracing \cite{lin6GDigitalTwin2023} and generative adversarial networks (GANs) \cite{10.5555/2969033.2969125, 10.1109/MCOM.2019.1800635}, this study assumes that statistical models such as those defined in 3GPP TS38.901 \cite{3GPP:38.901} strike a practical balance between accuracy and computational cost. Nonetheless, relying solely on simulations for training and validating these AI-enhanced receivers raises concerns about their real-world generalization. AI-assisted radios risk becoming biased toward the simulated environment and may perform poorly in actual deployments -- even if they excel in simulation. Despite growing interest in AI-based receivers, the literature on their over-the-air (OTA) performance remains scarce. Moreover, best practices for training these receivers are underexplored, revealing a critical gap.

To address this gap, we explore training strategies for AI-based receivers and evaluate their OTA performance. Specifically, we examine the impact of using individual 3GPP TS 38.901 channel models for training NN-based receivers. Furthermore, we investigate the effect of randomizing these channel models during training. We then validate these approaches through OTA measurements by constructing a complete orthogonal frequency-division multiplexing (OFDM) system \cite{changSynthesisBandLimitedOrthogonal1966} with software-defined radios (SDRs), and conduct OTA data collection in a controlled environment resembling a small-cell deployment. We compare the performance of these differently trained NN-based receivers against a conventional OFDM receiver employing Least Squares (LS) channel estimation and Linear Minimum Mean Square Error (LMMSE) equalization, chosen for its well-known performance-complexity trade-off and mathematical tractability. 

The potential of neural networks (NNs) for OFDM channel estimation has been demonstrated in studies such as \cite{7098027,8052521}, while other research has enhanced OFDM demodulation using NNs \cite{9882269}. Broader use cases and opportunities for NNs in this domain have also been identified in works like \cite{huangDeepLearningPhysicalLayer2020a,osheaIntroductionDeepLearning2017a} and beyond. In this study, we focus on DeepRx, a convolutional neural network (CNN)-based OFDM receiver that has significantly outperformed conventional LS/LMMSE receivers in simulations \cite{honkalaDeepRxFullyConvolutional2021}. DeepRx employs a fully convolutional architecture inspired by ResNet \cite{heDeepResidualLearning2016}, integrating channel estimation, equalization, and demodulation into a single CNN framework. We selected DeepRx for our research due to its holistic approach and robust performance in simulated environments.

This research evaluates a DeepRx receiver trained under various channel models and demonstrates generalization potential and real-world performance when trained with sufficiently rich data. It also stresses the necessity of diverse, randomized simulation data for model training and the critical role of OTA measurements in validating AI-assisted receivers. While our study focuses on a small-cell environment, the findings suggest that more complex scenarios warrant further investigation, including the use of OTA data for fine-tuning the receiver models, MIMO configurations, and higher center-frequencies in more complex environments. The main contribution of this work is to show that randomizing TS38.901 channel models, even with parameterizations vastly deviating from the measured radio environment, can diversify training data and yield robust OTA performance, thereby offering insights into effective training strategies for AI-based receivers. 

The remainder of this paper is organized as follows: Section~\ref{sec:methods} outlines the research methodology and system model, and the measurement results are then reported in Section~\ref{sec:results}. Section~\ref{sec:discussion} discusses the key observations, and finally, the conclusions are drawn in Section~\ref{sec:conc}.

\section{Methodology}
\label{sec:methods}

\subsection{OFDM System Model}

\begin{figure}[hbt]

    \begin{flushleft}
    \includegraphics[width=0.485\textwidth]{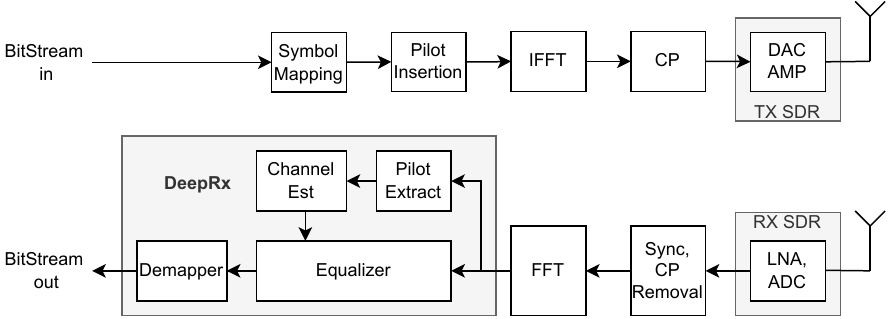}
    \caption{The implemented OFDM processing chain and SDR transceivers.}
    \label{fig:DeepRX_SDR}
    \end{flushleft}
\end{figure}

The block diagram of the implemented OFDM processing chain is depicted in Figure \ref{fig:DeepRX_SDR}. Aside from incorporating SDR for OTA measurements, the system model architecture closely adheres to the one outlined in \cite{honkalaDeepRxFullyConvolutional2021}.

\subsubsection{Conventional Receiver}

The conventional receiver follows a standard OFDM processing chain after the FFT block, which includes LS channel estimation, LMMSE equalization, and symbol-to-soft-bit demapping. It was selected as the benchmark for OTA tests because of its recognized performance-complexity trade-off, simple mathematical structure, and its extensive use and validation in existing wireless communication studies.

In particular, the received signal \( \mathbf{y}_{ij} \) over the $i$-th subcarrier and $j$-th OFDM symbol is modeled as:
\begin{equation}
    \mathbf{y}_{ij} = \mathbf{H}_{ij} \mathbf{x}_{ij} + \mathbf{n}_{ij}
\end{equation}
where \( \mathbf{y}_{ij} \in \mathbb{C}^{N_r \times 1} \) is the received signal, \( \mathbf{x}_{ij} \in \mathbb{C}^{N_t \times 1} \) is the transmitted signal, \( \mathbf{H}_{ij} \in \mathbb{C}^{N_r \times N_t} \) is the channel response, and \( \mathbf{n}_{ij} \in \mathbb{C}^{N_r \times 1} \) is the additive noise. Moreover, \( N_r \) and \( N_t \) denote the numbers of receive and transmit antennas, respectively.

The LS channel estimation measured from pilot signals is calculated as
\enlargethispage{-0.1in}
\begin{equation}
    \widehat{\mathbf{H}}_{p,\text{LS}} = \mathbf{y}_p \mathbf{x}^{H}_p
\end{equation}
where \( \mathbf{x}_p \) is an array consisting of the transmitted pilot symbols of all spatial streams for pilot index $p$ and $^{H}$ denotes Hermitian transpose. The channel is then interpolated between the pilots, which yields the complete channel estimate across all resource elements as $\widehat{\mathbf{H}}_{ij} = f \left( \hat{\mathbf{H}}_{p,\text{LS}} \right)$, where $f(\cdot)$ denotes the interpolation function.

The LMMSE equalized symbols are then calculated as
\begin{equation}
    \widehat{\mathbf{x}}_{ij} = \left( \widehat{\mathbf{H}}_{ij}^{H}  \widehat{\mathbf{H}}_{ij}  + \sigma^2 \mathbf{I} \right)^{-1} \widehat{\mathbf{H}}_{ij}^{H}  \mathbf{y}_{ij}
\end{equation}
where \( \sigma^2 \) is the estimated noise variance and \( \mathbf{I} \) is the identity matrix.

After this, the demapper maps the equalized symbols to Log-Likelihood Ratios (LLRs), which represent soft estimates of the received bits. For each received symbol \( \hat{x}_{ijk} \) of the equalized symbol vector $\widehat{\mathbf{x}}_{ij}$, the LLR of the \( k \)-th bit is given by:
\begin{equation}
    LLR_{ijk} = \log \left( \frac{\Pr(b_{ijk} = 0 \mid \hat{x}_{ijk})}{\Pr(b_{ijk} = 1 \mid \hat{x}_{ijk})} \right)
\end{equation}
where \( b_{ijk} \) denotes the \( k \)-th bit of the corresponding symbol.

\subsubsection{ML Receiver}

In our NN-based implementation, in contrast to the above-defined conventional architecture, the processing blocks in the receiver are replaced by a fully convolutional neural network, for which we employ ResNet structure \cite{heDeepResidualLearning2016}. The input to the NN is the received IQ signal for the whole TTI, and the output consists of the LLRs for all bits in the TTI. The NN is trained by optimizing cross-entropy loss between all ground truth bits and output LLRs. For further details about the NN-based DeepRx receiver, we refer the reader to \cite{honkalaDeepRxFullyConvolutional2021}.

\subsection{Numerology and Parameters}

The choice of operating frequency for our OFDM system was guided by license availability. Hence, despite common usage scenarios favoring higher frequencies, we settled on 434 MHz for compliance with regulations. The rest of the parameters used in our experiments are summarized in Table \ref{tab:parameters_table}.

\begin{table}[t]
\vspace*{0.1in} 
    \caption{Parameters}
    \begin{center}
    \begin{tabular}{|c|c|}
        \hline
        \multicolumn{2}{|c|}{\textbf{Measurement-only}} \\  
        \hline
        Center frequency & 433.92MHz \\
        \hline
        Bandwidth & 1.55MHz \\
        \hline
        Sampling resolution & 12 bits \\
        \hline
        SDR transceivers & AD9361-based \\
        \hline
        Antenna gains [TX, RX] & [10 dB, 0 dB] \\
        \hline
        Peak TX Power & 5 dBm \\
        \hline
        Mean TX PSD & -13 dBm / 100kHz \\
        \hline
        Decimation Factor & 16 \\
        \hline
        \multicolumn{2}{|c|}{\textbf{Simulation-only}} \\
        \hline
        Simulation environment & NVIDIA Sionna  \cite{hoydisSionnaOpenSourceLibrary2022}\\
        \hline
        SNR range & 10 dB to 35 dB \\ 
        \hline
        Channel models & All TDL and CDL, UMa, UMi\\
        \hline
        Speed range & 0 m/s to 30 m/s \\
        \hline
        Delay spread range & 50ns to 1000ns \\
        \hline
        \multicolumn{2}{|c|}{\textbf{Common}} \\
        \hline
        Modulation & OFDM 64QAM \\
        \hline
        FFT size [TX, RX] & [128, 128] \\
        \hline
        Subcarriers & 100 \\
        \hline
        Subcarrier spacing & 15kHz \\
        \hline
        Cyclic prefix & 6 \\
        \hline
        Pilot configuration & Every 4th SC, in 2nd symbol \\
        \hline
        Synchronization & Maximum preamble correlation \\
        \hline
        System & 1T1R SISO\\
        \hline
    \end{tabular}
    \label{tab:parameters_table}
\end{center}
\end{table}

\subsection{Radio Channel Simulation Algorithms}

For DeepRx training, we utilized various 3GPP statistical radio channel models \cite{3GPP:38.901}. Additionally, we investigated a randomized model to enhance the variability of simulated radio channels by randomly selecting one of the TDL or CDL model variants (A through E) during each training iteration. Unless otherwise specified, the simulated speed and delay ranges are set as detailed in Table \ref{tab:parameters_table}.


\subsection{Use of Software-Defined Radio (SDR) Device}

\begin{figure}[htbp]

    \includegraphics[width=0.46\textwidth]{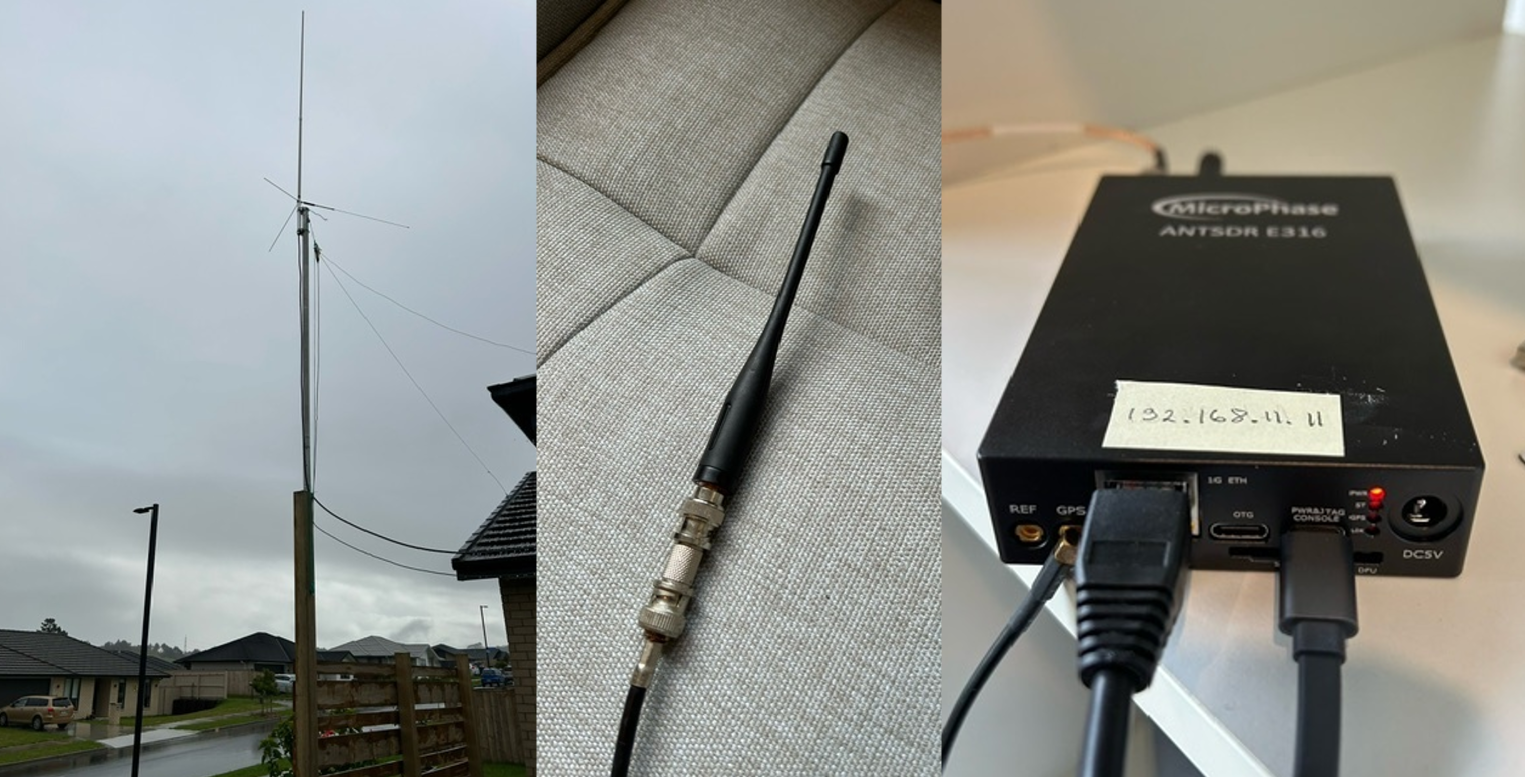}
    \caption{The transmitting antenna, the receiving antenna, and one of the two SDR radios}
    \label{fig:SDR}
\end{figure}

The OTA radio transmission tests were conducted using two SDRs. On the receive side, an omnidirectional antenna was connected to a 30 dB Low Noise Amplifier (LNA), which in turn was linked to the SDR’s RX port through a 15 m coaxial cable. The transmitting antenna was placed on a six-meter-high tower. Figure \ref{fig:SDR} depicts the setup, which emulates a small-cell environment with an outdoor base station (BTS) and an indoor UE.

The system used a preamble for synchronization, with a Zadoff-Chu sequence of length 100. GPS-disciplined OCXOs maintained frequency accuracy and stability, while decimation improved timing accuracy and reception.

\subsection{SDR Radio Channel Dataset Creation}

\subsubsection{SDR Data Collection}

Two data collection methods were utilized: one involved walking through each room of a steel-roofed, single-story building, and the other involved running through the same rooms while swaying the antenna in the air to increase time- and frequency-domain variability in the received OFDM signal. Collected data consisted of the originally transmitted bit stream, received QAM IQ symbols after synchronization and DFT, noise power, and SINR measurements per TTI. No significant external interference was detected during data collection.

\subsubsection{SDR Datasets}

Three datasets were constructed: a validation dataset of 500 TTIs collected while walking, a \textit{Test Dataset A} of 12,000 TTIs collected while walking, and a \textit{Test Dataset B} of 3,000 TTIs collected while running and swaying the antenna in the air for increased channel variability.

\subsection{ML model training Procedure}
\label{subsec:trainproc}

Training parameters are summarized in Table \ref{tab:training_params}. Each training sample was generated independently, and no validation or test samples were re-used in the training, which means that the model cannot overfit but may still be susceptible to distribution mismatch between training and validation. Training was manually stopped when no further improvement was observed, typically after around 50,000 iterations, or up to 150,000 iterations for models trained with randomized TDL/CDL channel models. Apart from the parameters described, the training procedure followed the guidelines outlined in \cite{honkalaDeepRxFullyConvolutional2021}.

\begin{table}[t]
    \vspace*{0.1in} 
    \caption{Training Parameters and Environment}
    \begin{center}
    \begin{tabular}{|{c}|p{5cm}|}
        \hline
        Batch Size & 28 \\
        \hline
        Initialization & He normal \\
        \hline
        Optimizer & AdamW, weight decay $1\times 10^{-3}$ \\
        \hline
        Loss Function & EbNo weighted Binary Cross-Entropy (BCE) \\
        \hline
        Learning rate & Exponential. Start $4\times 10^{-4}$, decay rate 0.6 every 8000 iterations. Minimum learning rate is $2\times 10^{-5}$ \\
        \hline
        Training dataset & Created on the fly, infinite.\\
        \hline
        Validation dataset & SDR generated, walking, 500 TTIs.\\
        \hline
        Test Dataset A & SDR generated, walking, 12000 TTIs\\
        \hline
        Test Dataset B & SDR generated, running, 3000 TTIs \\
        \hline
        RMS delay spread & 105ns, mean\\
        \hline
    \end{tabular}
    \label{tab:training_params}
\end{center}
\end{table}

\subsection{Performance Evaluation Criteria}

During training, the BCE loss was monitored for both simulated and measured validation data. The final model's performance was evaluated by comparing the Bit Error Rate (BER) as a function of SINR, with LS/LMMSE performance used as a benchmark.

\section{Results}
\label{sec:results}

In this section we present the OTA performance results of DeepRx receivers trained with various 3GPP TS38.901 channel models and parameter settings, including the receiver trained with randomized models. We also investigate the impact of delay spread and UE speed channel simulation parameters on DeepRx performance. The performance was measured with \textit{Test Dataset A} and \textit{Test Dataset B}, which were generated OTA using an SDR. 

\subsection{Channel Models}

First, we examine the performance of DeepRx receivers trained with individual channel models, evaluated on \textit{Test Dataset A}, which was collected at walking speed. 

\subsubsection{3GPP TS38.901 TDL, UMa and UMi channel models}
\label{subsec:tdl_um}

\begin{figure}[htbp]

    \includegraphics[width=0.47\textwidth]{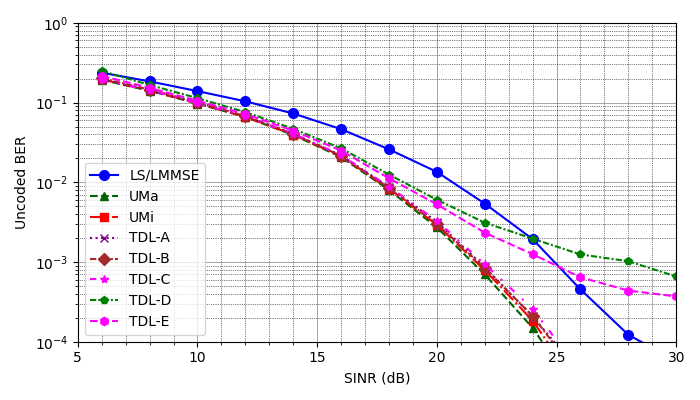}
    \caption{Uncoded BER versus SINR on SDR \textit{Test Dataset A} for DeepRx models trained with TDL, UMa, and UMi channel models.}
    \label{fig:results_TDL}
\end{figure}

Figure \ref{fig:results_TDL} illustrates the OTA results of DeepRx receivers trained with TDL-A through TDL-E channel models, as well as UMa and UMi models. During training, the simulated speed ranged randomly from 0 to 30 m/s, while the delay spread varied from 50 ns to 1 µs, i.e. the parameters shown in Table \ref{tab:parameters_table}. 

Notably, LS/LMMSE receiver is outperformed by DeepRx trained with every channel model except TDL-D and TDL-E, which perform worse at higher SINRs. Both UMa and UMi models, however, perform well.

\subsubsection{3GPP TS38.901 CDL and the randomly selected channel models}
\label{subsec:cdl_random}

\begin{figure}[htbp]

    \includegraphics[width=0.47\textwidth]{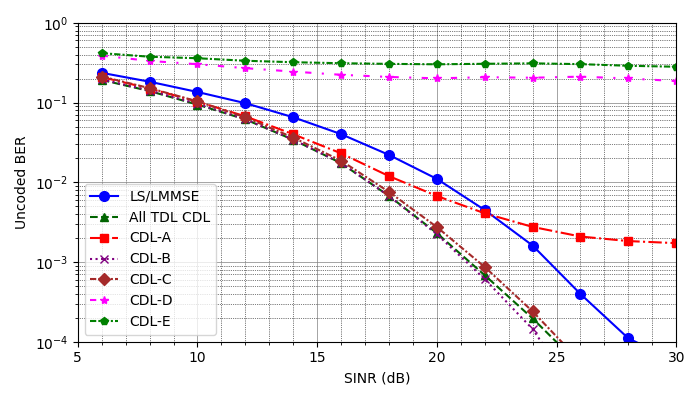}
    \caption{Uncoded BER versus SINR on SDR \textit{Test Dataset A} for DeepRx models trained with CDL and with \textit{ALL TDL/CDL}, a randomly generated mixture of TDL and CDL models.}
    \label{fig:results_CDL}
\end{figure}

Figure \ref{fig:results_CDL} shows the OTA performance results of DeepRx models trained with CDL-A through CDL-E alongside a model trained with randomly selected TDL and CDL variants, for each training iteration. This randomly generated model is labeled \textit{ALL TDL/CDL} in the figures. For consistency, we employed the same speed and delay spread ranges as in subsection \ref{subsec:tdl_um} during training.

While CDL-B and CDL-C demonstrate substantial performance improvements, CDL-A is underperforming. Although CDL-D and CDL-E, which represent line-of-sight (LOS) channels, seemed to converge during training, they perform poorly in the OTA tests. The randomized \textit{ALL TDL/CDL} model equally well with CDL-B and CDL-C.

\subsection{Delay Spread and Speed}

We next explore the influence of model parameters on OTA performance. Specifically, we evaluate DeepRx trained with varying delay spread and speed ranges to understand how these variables impact its effectiveness. For these parameter impact tests, we chose TDL-B as the base model because of its relatively strong performance when validating against OTA data in our experimental environment.

\subsubsection{Delay Spread}

\begin{figure}[htbp]

    \includegraphics[width=0.47\textwidth]{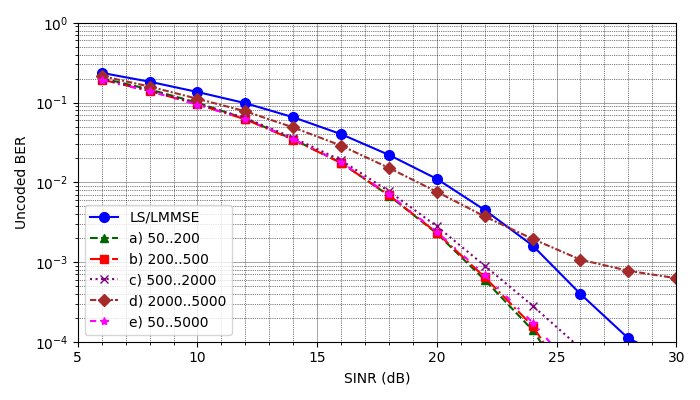}
    \caption{Uncoded BER versus SINR on SDR \textit{Test Dataset A} for DeepRx models trained with TDL-B under different delay spread ranges. Units in the legend are nanoseconds.}
    \label{fig:results_SPREAD}
\end{figure}

Figure \ref{fig:results_SPREAD} presents the performance of the TDL-B model trained with varying delay spread ranges, evaluated on \textit{Test Dataset A}. Consistent with expectations, models trained with shorter delay spreads reflecting the conditions of the actual measurement environment outperform those with longer delay spreads. Importantly, a DeepRx model trained across an extensive delay spread range of 50 ns to 5 µs performs on par with the best-performing models trained with narrower delay spread ranges.

\subsubsection{Speed}

\begin{figure*}[htbp]

  \subfloat[Low speed]{\includegraphics[width=0.49\textwidth]{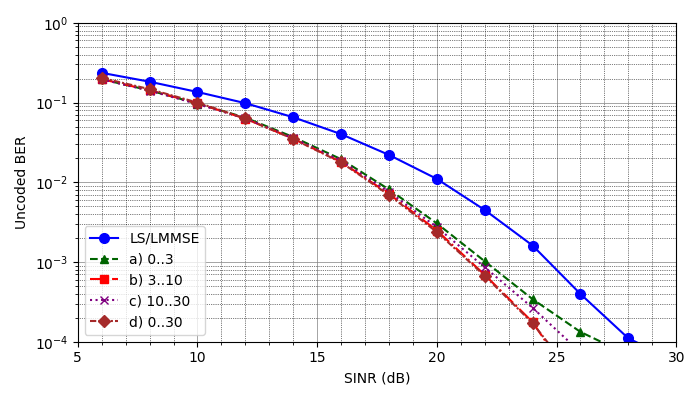}}
  \subfloat[High speed]{\includegraphics[width=0.49\textwidth]{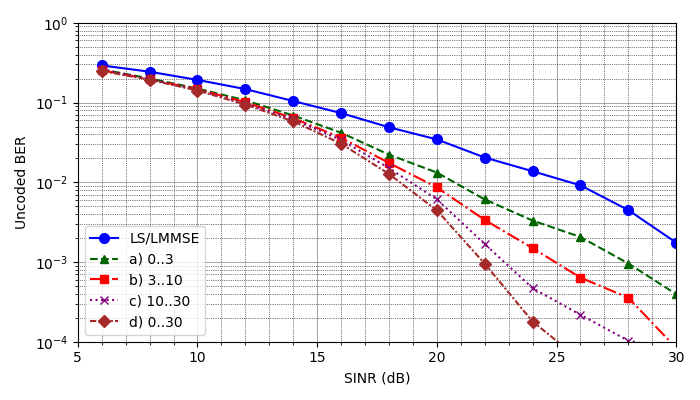}}
       \caption{Uncoded BER versus SINR on SDR \textit{Test Dataset A} (a) Low speed and \textit{Test Dataset B} (b) High speed for DeepRx models trained with TDL-B at various speed ranges. Units in the legend are m/s.}
    \label{fig:results_SPEED_LO}\label{fig:results_SPEED_HI}
\end{figure*}

Figure \ref{fig:results_SPEED_LO} illustrates the performance of DeepRx trained with TDL-B accross various simulated speed ranges. For \textit{Test Dataset A}, which was collected at walking speeds well below 3 m/s, contrary to expectations, DeepRx models trained with simulated speeds below 3 m/s  perform worse than those trained with higher speeds. The best results are achieved by training over a broad speed range of 0 to 30 m/s. 

A similar pattern can be observed for \textit{Test Dataset B}, which was collected while running and swaying the antenna in the air to increase time- and frequency-domain variability. Also in this scenario, a DeepRx model trained with a speed range of 0 to 30 m/s delivers the best performance, matching the performance of the model tested with \textit{Test Dataset A}. Again, models trained with speed ranges specifically aligned to the measurement conditions show reduced performance. Furthermore, the LS/LMMSE receiver's performance is significantly degraded under these elevated speed conditions.

Overall, results indicate that training with diverse, randomized simulations leads to robust generalization, while narrowly scoped, individual channel models can risk compromising both training and validation of the trained receiver.

\section{Discussion}
\label{sec:discussion}

The OTA test results show that DeepRx, a CNN-based receiver, can significantly outperform the LS/LMMSE receiver when trained with randomized 3GPP TS38.901 channel models. By randomly selecting channel models for each training iteration and applying a broad range of delay spreads and UE speeds, DeepRx demonstrated robust generalization to our test environment, matching or exceeding the performance of receivers trained with narrowly targeted channel models. 

Although training the receiver with randomized 3GPP TS38.901 channel models resulted in performance comparable to the best individual models, training DeepRx with individual models—despite incorporating broad delay spreads and UE speed settings—led to inconsistent outcomes; in some instances, performance barely exceeded random guessing. This indicates that individual 3GPP TS38.901 models may be too narrow or leave gaps in the spectrum of radio channel realizations.

To further investigate the effects of simulated parameters on model performance, we subdivided the speed and delay spread ranges into narrower segments. We used TDL-B for this analysis based on its solid performance in prior tests. Results show that the best performance was achieved when the delay spread closely matched the test environment, while performance gradually degraded with increasing simulated delay spreads. Notably, models trained across a wide range of delay spreads \begin{math} \left(50 \text{ ns to } 5\, \mu\text{s}\right) \end{math} performed similarly to those trained on a narrower range resembling the test environment—confirming that training DeepRx with broad parameter ranges supports real-world generalization. 

Counterintuitively, DeepRx performed worse when trained with simulated speeds matching the actual measurement speed \begin{math}\left(<3\, \text{m/s}\right)\end{math} compared to when trained with higher speeds \begin{math}\left(10 \text{ m/s to 30 m/s}\right)\end{math}. This discrepancy became more pronounced when test data were collected by running while swaying the antenna in the air (\textit{Test Dataset B}), thereby introducing greater channel variability. These findings suggest that the simulation may not accurately capture the test environment or may omit certain channel conditions. By contrast, higher simulated speeds appeared to generate a broader spectrum of channels, enhancing DeepRx’s learned features and, thus, its overall performance. This indicates that highly randomized simulations optimized for NN training—rather than those attempting to mirror real-world conditions—could be more effective, presenting a promising direction for future research to expand the applicability of learned receivers in practical deployments.

While these results highlight DeepRx’s promising generalization capabilities, the study does have limitations. Our measurements were restricted to a relatively narrow environment, and broader testing—including more complex environments and higher carrier frequencies—would offer deeper insights into NN-based receivers’ generalization. It is also worth noting that while DeepRx was chosen for its demonstrated performance in prior simulations, other NN-based receiver architectures might generalize differently in real-world conditions. Nevertheless, these observations underscore the importance of using broad and varied training data, and the necessity of OTA testing. Specifically tailored simulations for training NN-based receivers could significantly enhance their performance and applicability in real-world scenarios. Future research should also explore the benefits of fine-tuning receiver models with OTA-collected data.

\section{Conclusion}
\label{sec:conc}

\balance

While radio channel simulations can provide a practical data source for training NN-based receivers, relying solely on them for model validation introduces uncertainty regarding real-world performance. In this study, we investigated randomized 3GPP TS38.901 models for training DeepRx, a CNN-based OFDM receiver, and validated its performance with OTA experiments in a small cell-type setting. Our results showed that DeepRx trained with broad, diverse channel data can significantly outperform conventional LS/LMMSE receivers, whereas training with individual 3GPP TS38.901 models produced varying outcomes. In particular, it was observed that simulations aiming to mirror real-world conditions may not be optimal for training AI receiver models. This highlights the necessity for effective training strategies and customized radio channel models to develop and train robust AI-based receivers. Beyond tailored channel models, future efforts should explore the use of OTA data for fine-tuning, alternative receiver architectures, incorporate MIMO configurations, and investigate higher carrier frequencies in more complex environments.

\bibliographystyle{ieeetr}
\bibliography{main}

\end{document}